\documentclass[12pt]{article}

\usepackage{times}

\usepackage{bm}

\usepackage{ulem}
\usepackage{graphicx}
\usepackage{amsfonts}

\usepackage{amsmath, amsthm, amssymb}

\usepackage{color}

\usepackage{url}
\usepackage{scicite}
\usepackage{graphicx}
\usepackage{url}

\usepackage{dcolumn}

\usepackage{color}

\newcommand{\beq}{\begin{equation}}
\newcommand{\eeq}{\end{equation}}

\newcommand{\pd}{\partial}

\newcommand{\ybal}{$\beta$-YbAlB$_4\,$}
\newcommand{\ybalfe}{$\beta$-YbAl$_{1-x}$Fe$_x$B$_4\,$}

\newenvironment{sciabstract}{%
\begin{quote} \bf}
{\end{quote}}

\newcounter{lastnote}

\title{Strange Metal Without Magnetic Criticality}

\author
{Takahiro Tomita$^{1,2}$, Kentaro Kuga$^1$, Yoshiya Uwatoko$^1$,\\
Piers Coleman$^{3,4}$ and Satoru Nakatsuji$^{1,5,\ast}$\\
\normalsize{$^{1}$Institute for Solid State Physics, University of Tokyo, Kashiwa 277-8581, Japan,}\\
\normalsize{$^{2}$College of Humanities \& Sciences, Nihon University, Setagaya 156-8550, Japan,}\\
\normalsize{$^{3}$Center for Materials Theory, Department of Physics and Astronomy,}\\
\normalsize{Rutgers University, Piscataway, N.J. 08854, USA}\\
\normalsize{$^{4}$Department of Physics, Royal Holloway, University
of London, Egham, Surrey TW20 0EX, UK.}\\
\normalsize{$^{5}$PRESTO, Japan Science and Technology Agency (JST),}\\
\normalsize{4-1-8 Honcho Kawaguchi, Saitama 332-0012, Japan}\\
\normalsize{$^\ast$To whom correspondence should be addressed; E-mail:  satoru@issp.u-tokyo.ac.jp.}
}

\date{}

\begin{document} 


\baselineskip24pt


\maketitle 
\newpage


\begin{sciabstract}
A fundamental challenge to our current understanding of metals is the
frequent observation of qualitative departures from Fermi
liquid behavior. The standard view attributes such non-Fermi liquid phenomena
 to the scattering of electrons off quantum critical fluctuations
of an underlying order parameter. While the possibility of non-Fermi liquid behavior 
isolated from the border of magnetism has long been speculated, no experimental confirmation has been made. Here we
report on the observation of a strange metal region in the absence of a
magnetic instability in an ultrapure single crystal. In particular, we
show that the heavy fermion superconductor \ybal forms a possible phase with
strange metallic behavior across an extensive pressure regime,
distinctly separated from a high-pressure magnetic quantum phase
transition by a Fermi liquid phase.

\end{sciabstract}

\newpage



Extensive investigations of strongly correlated materials over
past decades have demonstrated that qualitative 
deviations from the standard theory of metals, 
Landau's Fermi liquid (FL) theory \cite{LandauFL}, develop
almost routinely in the vicinity of a magnetic quantum phase
transition \cite{lohneysen2007,gegenwart08}.
Conventionally, the origin of such non-Fermi liquid (NFL) behavior
is attributed to the strong damping of the quasi-particle's
life time by quantum critical fluctuations of an underlying order
parameter
\cite{hertz-76,Moriya85,millis-93,Coleman01,Si01,grigera-01}. 

Physics delineates  between the
concept of a phase, occupying a finite parameter region of
ground-state, and quantum critical points, appearing
at the transition between phases. Although the possible 
existence of strange metal
phases with NFL behavior, 
occupying a finite region of the ground-state phase diagram
has long been speculated
\cite{Senthil03,MnSi,Canfield,si2006,takashima2007,anderson2009,Mihailo2009,Friedemann2009,Thomas2010,Custers2010},
the close proximity of such phenomena to magnetic
instability, and a strong sensitivity to impurities has to date
impeded confirmation of this idea. One of the most challenging questions is whether a fully paramagnetic strange metal phase is possible without magnetic criticality, retaining full symmetry of the underlying crystal structure. 

Many prototypical quantum critical (QC) materials have been found within the class of
$4f$ heavy fermion compounds.  The highly tunable characteristic
energy scales and availability of high purity crystals make them ideal
candidates for the study of quantum
criticality\cite{lohneysen2007,gegenwart08}.  In these
materials, quantum criticality develops from a competition between
local moment magnetism and the conduction electron screening of the
local moments (the Kondo effect).  Most QC heavy-fermion materials are
known to have an almost integral valence which stabilizes the local
moments considered essential for the criticality.

An exception to this rule was recently discovered in \ybal, which
exhibits quantum criticality despite strong mixed valency\cite{Review6, ybal-valency,Matsumoto2011Science, MachidaSeebeck}.
Ultrapure single crystals of this material exhibit intrinsically singular thermodynamic and transport
behavior up to an upper limit scale of several K, including a divergent temperature dependence of the magnetic susceptibility $\sim
T^{-1/2}$ and an anomalous  $T^{3/2}$ dependence of the electrical resistivity, both of 
which are extremely sensitive to a magnetic field $B$ \cite{Review6,Matsumoto2011Science,MachidaSeebeck}.
In particular, $T/B$
scaling of the magnetization has been observed over four decades of
$T/B$, projected to extend down to fields as small as 0.1 mT
\cite{Matsumoto2011Science}.
However, the observation of intrinsic quantum criticality
as a function of field  does not rule out the possibility that this
phenomenon is merely a fine-tuned coincidence of lattice structure.
Here through an extensive series of pressure measurements of the resistivity using ultrapure crystals,
we demonstrate that the intrinsic quantum criticality of \ybal is not fine-tuned, but instead occupies an extended island of pressure in the phase diagram, indicating a formation of a phase without any symmetry breaking external fields for stabilization. 
Furthermore, we show that the strange metal region is clearly surrounded and 
separated from a high-pressure magnetic instability by
a finite pressure range of Fermi liquid behavior. 

\begin{figure}
\begin{center}
\hspace{0cm} 
\includegraphics[width=7in]{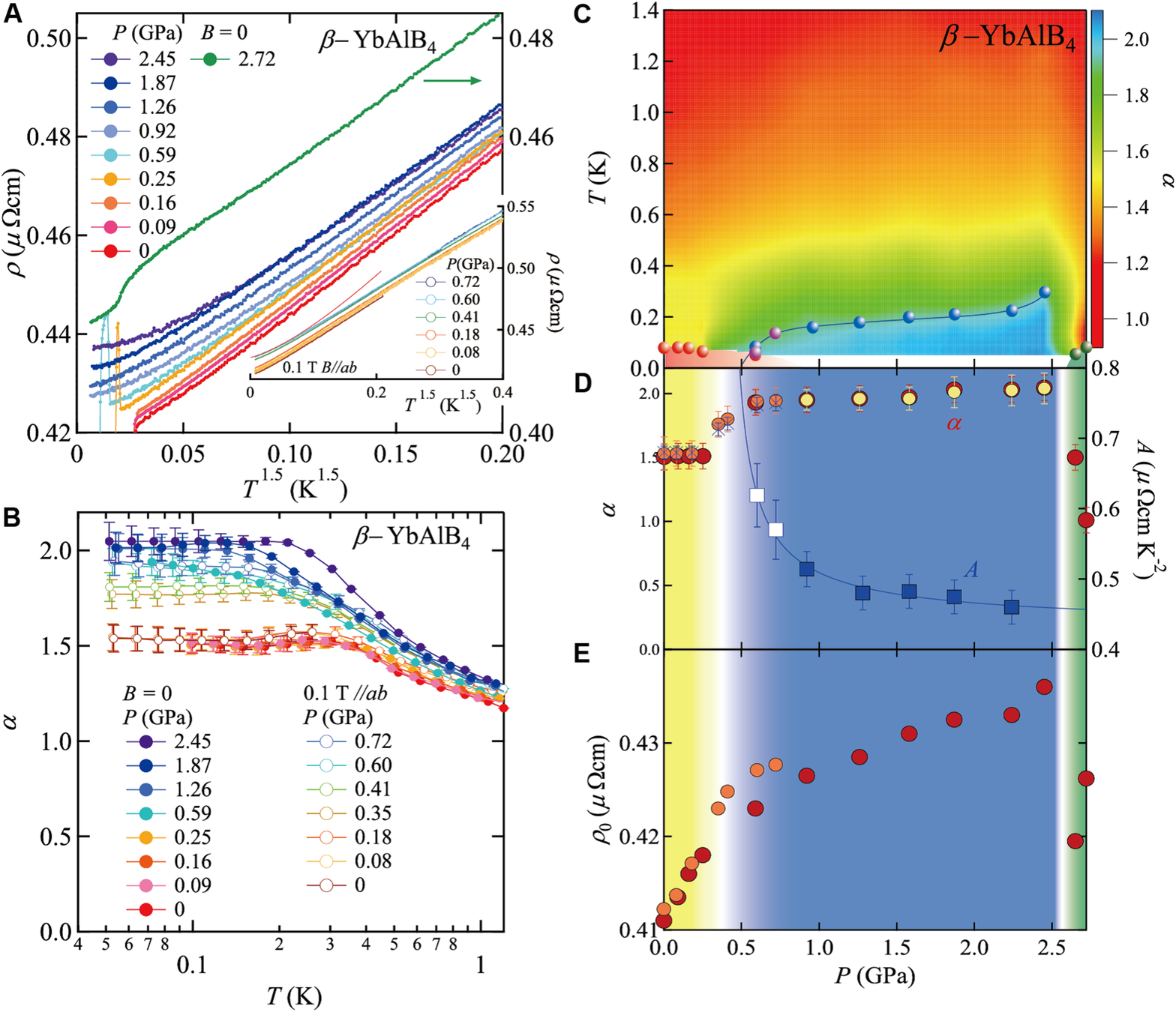} 
\caption {
\label{PD} }
\end{center}
\end{figure}

First we present our main experimental observation of the extensive region of the strange metal behavior and its evolution to a Fermi liquid phase. 
Figure \ref{PD}{\bf A} shows the  temperature dependence of the zero-field resistivity of \ybal under various pressures. Here, we employed ultrapure crystals with RRR $= 300$  (with $\rho_0 < 0.5~\mu \Omega$cm, and mean free path of $> 1000$ \AA ) and performed high-precision resistivity measurements (with noise levels of $< 40$ pVHz$^{-1/2}$) using a piston-cylinder pressure cell in a dilution refrigerator \cite{SOM}. The pressure was continuously monitored using tin and aluminium superconducting manometers.  
X-ray diffraction analyses confirm a continuous reduction of the lattice parameters under pressure with a bulk modulus of 189 GPa (fig. S\ref{S3}) \cite{SOM}.
Strikingly, under pressures up to 0.25 GPa, the resistivity exhibits the same anomalous power law behavior $\rho(T)\sim T^{1.5}$ with the same slope as at ambient pressure. In contrast, above $P_{\rm c} \sim 0.4$ GPa, $\rho(T)$ shows a clear deviation from a $T^{1.5}$ dependence at low $T$s and exhibits a FL-like $T^2$ dependence. This can be clearly seen in the $T$ dependence of the power law exponent $\alpha$, (solid circle) in $\rho(T) -\rho_0 \propto T^\alpha$, as shown in Fig. \ref{PD}{\bf B}. Interestingly,  under $P < P_{\rm c}\sim 0.4$ GPa,  $\alpha$ increases gradually on cooling and becomes constant $\sim 1.5$ below 0.3 K down to the superconducting (SC) transition temperature  $T_{\rm c}$. In contrast, at $P > 0.4$ GPa it saturates to $\alpha  = 2.0$, the value known for a FL state. 

The superconducting $T_{\rm c}$ continuously decreases with pressure from $T_{\rm c} = 80$ mK at ambient pressure,  and finally vanishes around 0.6 GPa (Figs. \ref{PD}{\bf A \& C}). To extend our analysis below $T_{\rm c}$,  we measured the  resistivity  by suppressing SC under a weak magnetic field along the $ab$-plane, which should be irrelevant to the NFL critical fluctuations due to the Ising character of the $4f$ moments.  Figure  \ref{PD}{\bf A}  inset plots the resistivity vs. $T^{1.5}$ measured at an in-plane field $B_{ab} = 0.1$ T  under various pressures.  
The corresponding $T$ dependence of the exponent (shown in Fig. \ref{PD}{\bf B}) indicates that 
 the strange metallic state with $\alpha = 1.5$ extends down to the lowest $T \sim 50$ mK under $ P < P_{\rm c}$, while the exponent saturates to $\alpha \sim 1.8$ at $P \sim P_{\rm c}$ and to $\alpha = 2$  for $P > P_{\rm c}$. The Fermi liquid temperature $T_{\rm FL}$, below which $\rho(T)$ shows $T^2$ law, systematically decreases with decreasing pressure and appears to vanish at $ P \sim P_{\rm c}$ (fig. S8).


Figure \ref{PD}{\bf C} provides the contour plots of the exponent $\alpha$ obtained using the zero-field $\rho (T)$ data in Fig. \ref{PD}{\bf A}. The diagram reveals an extended region with anomalous exponent  indicating the formation of the strange metal phase, and its subsequent crossover into the high pressure FL phase. 
From ambient pressure, a NFL (yellow) region with $\alpha = 1.5$ occupies a
finite range up to $P_{\rm c} \sim 0.4$ GPa above the SC dome. 
In contrast, at pressures beyond $P_{\rm c}$ up to 2.5 GPa, $\alpha$ locks into a constant $\sim 2$ (blue) below $\sim 100$ mK, indicating the formation of a FL phase.
To carefully examine the phase evolution,  in Fig. \ref{PD}{\bf D} we plot the exponents $\alpha$= $\partial \ln (\rho(T)-\rho_0)/\partial \ln T$ at the midpoints of two temperature ranges: $90\sim 120$ mK  under zero field  (large red circles), and $40\sim 60$ mK under an in-plane field of 0.1 T to suppress the SC (orange circles). For the non-SC region at $P > 0.8$ GPa, we plot the zero-field  exponent for the $T$ range, $40\sim 60$ mK (yellow circles). To further evaluate the exponent without the ambiguity associated with the residual resistivity, we have also carried out the analysis of the resistivity exponent  (cross) using $\alpha=1+ \partial \ln (\partial \rho(T)/\partial T)/\partial \ln T$  in the $T$ range $40 \sim 80$ mK, by suppressing the SC under $0.1$ T at $P < 0.8$ GPa.
All the data are consistent with the existence of a NFL phase with a constant $\alpha \approx 1.5$ at $ P < P_{\rm c}$  and a FL phase with $\alpha = 2.0$ at  $P > P_{\rm c}$ (Fig. \ref{PD}{\bf D}). The apparent crossover between $\alpha = 1.5$ and 2.0 marked by the two points with intermediate exponents is most likely a consequence of experimental resolution and a small inhomogeneity in the pressure. The presence of the superconducting resistivity spike close to $P_{\rm c}$ (Fig. \ref{PD}{\bf A}) supports this interpretation \cite{SOM}.

In the FL phase, the $A$ coefficient for the $\rho \sim T^2$ law are found to be field-independent at $B_{ab} \le 0.1$ T.  Figure \ref{PD}{\bf D} shows that a part of the $A$ coefficient exhibits  a divergence at
$P_{\rm c}$, following $\sim 1/(P - P_{\rm c})^{\rm 0.8(1)}$ with $P_{\rm c} = 0.40(5)$ GPa.
In addition, kinks in $\rho_0$ are observed around $P_{\rm c}$ for both $B=0$ and 0.1 T (Fig. \ref{PD}{\bf E}). 
Taken together with the change in the exponent $\alpha$, these anomalies suggest a possible quantum phase transition at $P_{\rm c}$ separating the strange metal phase from the high pressure FL. 


Each of the putative NFL phases reported to date
directly adjoin a  magnetic phase and
are thus linked to 
 magnetic criticality
\cite{MnSi,Canfield,takashima2007,Friedemann2009,Custers2010}.
Generally, in Yb based heavy fermion compounds, both physical and
chemical pressure induce magnetism, stabilizing an ``Yb$^{3+}$'' state
with a $4f$ magnetic moment and a smaller ionic radius than its
nonmagnetic ``Yb$^{2+}$'' counterpart \cite{Goltsev}. 
To clarify the relation between magnetism and the observed extensive regime of NFL behavior in \ybal, we
have performed a detailed study using
high pressure and 
chemical substitution.

\begin{figure}
\begin{center}
\hspace{-15mm} 
\includegraphics[width=7in]{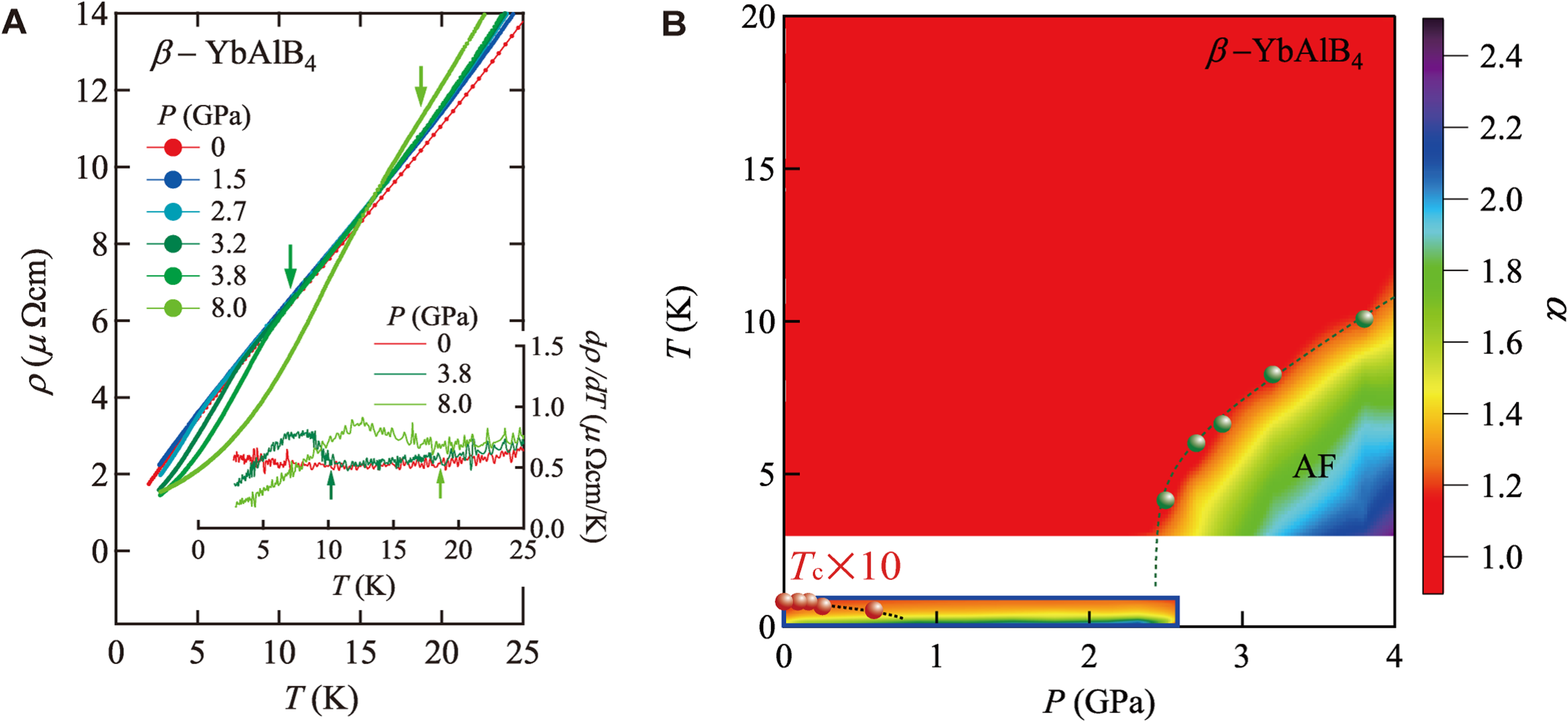} 
\caption {
\label{P-Fe}
}
\end{center}
\end{figure}

First, let us discuss the results of the ``high-$T (> 2$ K)" resistivity measurements performed in a cubic anvil cell that allows us to reach a much higher pressure, up to 8 GPa (Fig. \ref{P-Fe}{\bf A})\cite{SOM}. 
While a systematic change is found in the resistivity $\rho(T)$ at $T > 10$ K, no change was found in $\rho(T)$ at $P \le 2.3$ GPa below $\sim 10$ K \cite{SOM}. 
Figure \ref{P-Fe}{\bf B } shows the contour plots of the resistivity exponent $\alpha$.
By far the most prominent feature of the phase diagram is the wide (red) region of  anomalous $T$-linear resistivity \cite{SOM}. This region 
spans from ambient pressure to 3 GPa, extending over a decade of
$T$ from $\sim 2$ to 20 K (Figs. \ref{P-Fe}{\bf A} \& {\bf B} ).
Beyond the critical
pressure $P_{\rm N} \sim 2.5$ GPa, a kink develops in the resistivity,
where the temperature derivative $d\rho (T)/dT$ changes abruptly (Fig. \ref{P-Fe}{\bf A} inset). The
``kink'' temperature $T_{\rm N}$ marks the development of antiferromagnetic (AF) order, as we will discuss.
$T_{\rm N} (P)$ rises rapidly to 18 K at 8 GPa, to
our knowledge, the highest N\' eel point in Yb
based heavy fermion systems.

Correspondingly, in the ``low-$T (< 1$ K)" measurements using the dilution refrigerator, application of pressures exceeding $P_{\rm N} \sim 2.5$ GPa in a piston cylinder cell  gives rise to a sudden decrease in $\rho_0$ (Fig. \ref{PD}{\bf E}); moreover a kink develops in the resistivity and its $T$ derivative at a temperature $T_{\rm N}$, which rapidly rises from 80 mK at 2.72 GPa to $\sim 4$ K at 2.8 GPa (Fig. \ref{PD}{\bf A},  fig. S9). 
Within the pressure uncertainty, this coincides with the onset of antiferromagnetism found in the cubic anvil cell (fig. S6A inset).
This rapid increase of $T_{\rm N}$ as well as the jump in $\rho_0$ across  $P_{\rm N}$ suggests the pressure-induced magnetic phase transition is first order.

\begin{figure}
\begin{center}
\hspace{-1cm}
\includegraphics[width=7in]{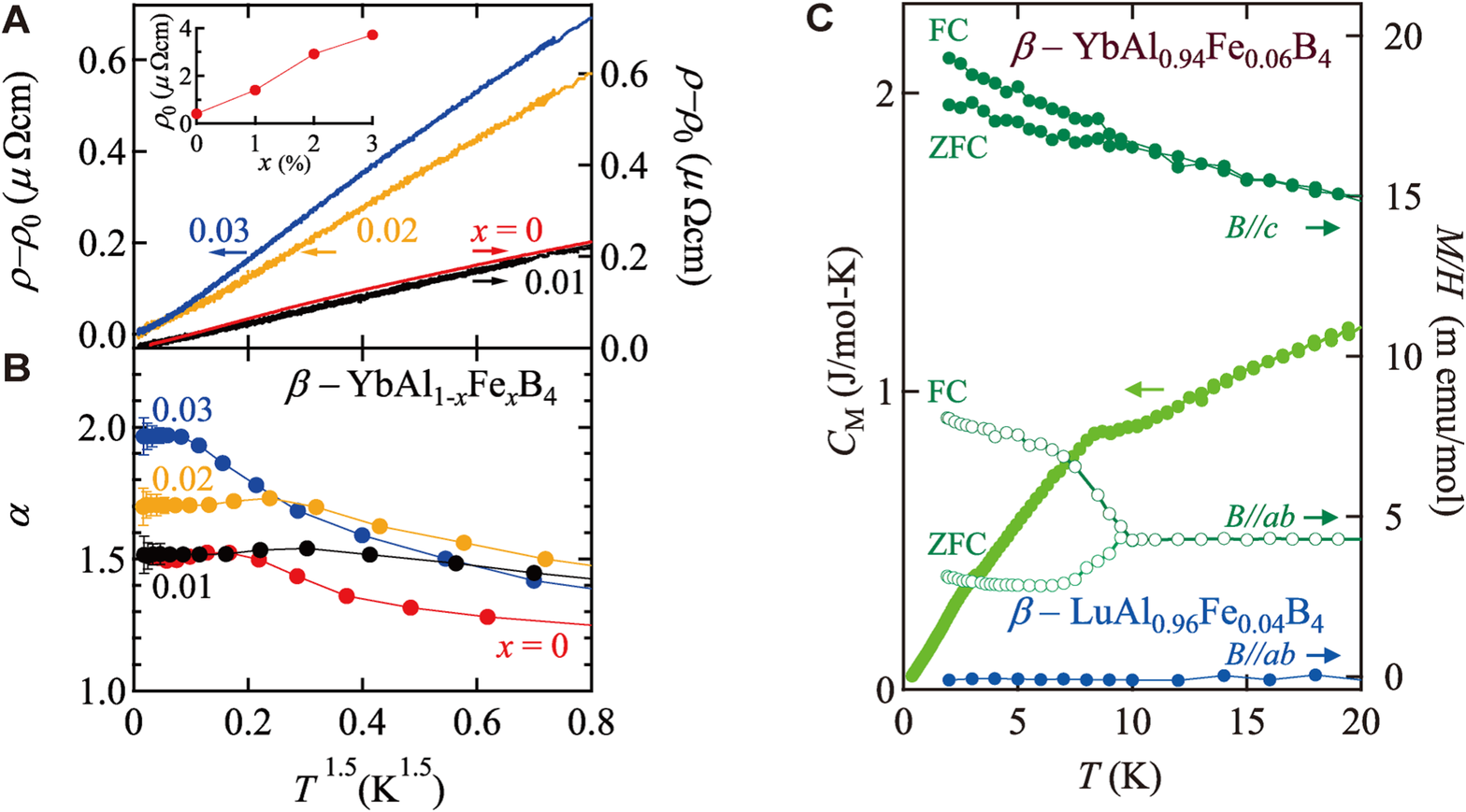}
\caption {\label{LowT-AF}
}
\end{center}
\end{figure}

Chemical substitution confirms a similar phase
evolution to 
that under pressure. 
In particular, Fe substitution for Al is found to lead to a crossover from a distinct region with quantum critical behavior to a Fermi liquid.
Figures \ref{LowT-AF}{\bf A}\&{\bf B} show the $T$ dependence of the
resistivity and its power law exponent, respectively.  The chemical analysis as well as
the systematic increase in $\rho_0$ confirms a homogeneous
distribution of Fe ions (Fig.  \ref{LowT-AF}{\bf A} inset) \cite{SOM}.  With 1 \% doping of Fe, we found
the power law exponent $\alpha$ in $\rho(T)$, approaches 1.5 upon
cooling below 1 K, the same anomalous exponent as in pure \ybal,
indicating the formation of the strange metal phase.  At higher Fe
content of $x = 2$ \% and 3 \%, on the other hand, the exponent
$\alpha$ approaches 1.7 and 2.0 respectively.  In addition, both
$\chi(T)$
and $C_{\rm M}(T)$ for $x = 3$ \% show no magnetic anomaly, but level
off on cooling, signaling the formation of a Fermi liquid
(fig. S\ref{S4})\cite{SOM}.  

Moreover, a 6 \% substitution of Fe contracts the volume by 0.6(2) \% and induces antiferromagnetism (Fig. \ref{LowT-AF}{\bf C}, table S1)\cite{KugaPRB,SOM}. The susceptibility $\chi(T)$ shows a kink at 9 K and a weak hysteresis between field-cooled and zero-field-cooled sequences, typically a 
signature of canted AF \cite{KugaPRB}.
The specific heat $C_{\rm M}(T)$ confirms the bulk nature of the magnetism,
showing an anomaly at 8.5 K.  By contrast, the Lu nonmagnetic analogue,
$\beta$-LuAl$_{1-x}$Fe$_x$B$_4$, exhibits diamagnetism. Thus, the magnetism derives
from the Yb rather than the Fe sites. 

\begin{figure}
\begin{center}
\hspace{0cm}
\includegraphics[width=6in]{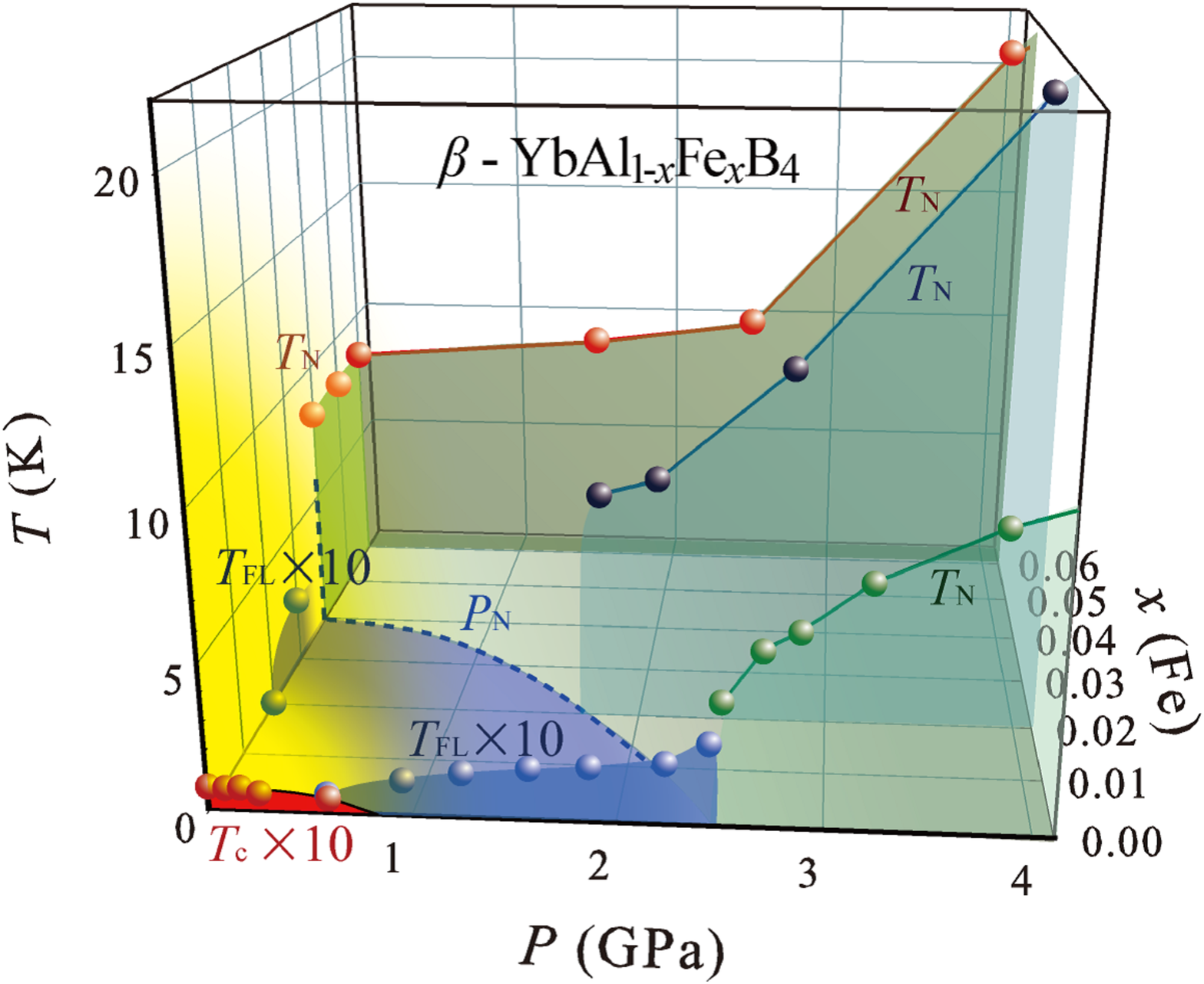}
\caption {\label{SPD}
 }
\end{center}
\end{figure}

Application of pressure to the 6 \% Fe substituted \ybal systematically increases the N\' eel temperature $T_{\rm N}$ up to 25 K at $\sim 5.5$ GPa (Fig. \ref{SPD}, fig. S\ref{S7}).  For $x = 2$ \% Fe substitution, pressure also induces magnetism at a critical
pressure $P_{\rm N} \sim 2$ GPa, a lower value than in the undoped crystals (2.5 GPa). Figure \ref{SPD} summarizes the combined data in a single phase diagram spanned by pressure ($P$),
Fe concentration ($x$), and temperature ($T$) axes.  
The smooth evolution of $T_{\rm N}$ as a function of pressure and doping strongly suggests that the pressure-induced phase in pure \ybal involves the same type of AF order found in the Fe doped \ybal.


Conventionally, quantum criticality develops at a zero temperature phase transition into a broken symmetry state. 
In \ybal, however, we find an intermediate FL phase nestled between the NFL region and the AF phase, showing that the NFL is not associated with the broken symmetry phase transition. 
This indicates that the origin of the low-pressure quantum criticality is a different kind of electronic instability.

One possibility is a topological phase transition. 
There are two such proposals that have been advanced in the literature. 
One is that the observed  criticality is
associated with the partial Mott
localization of the 
$f$-electrons to form a decoupled neutral spin liquid with 
fractionalized spin-1/2 excitations, co-existing
with a small-Fermi surface Fermi
liquid (FL$^{*}$) \cite{Senthil03,Senthil2004,Paul07} . 
In this scenario, as pressure is
applied to \ybal, the increased localization of the $4f$ electrons 
gives
rise  to  a spin liquid phase, stabilized by frustration in the honeycomb
layers and the presence of valence fluctuations \cite{si2006,Custers2010}. 
The observed quantum criticality
would arise as a gapless intermediate critical phase, 
screened by low $T$ SC\cite{Senthil03,Senthil2004}, which  
separates the
heavy FL with a large Fermi surface from 
a topologically distinct high-pressure
FL$^{*}$ with a small Fermi surface. 

An alternative possibility, is the formation of a vortex metal\cite{ramires2012beta}.  In \ybal, the
high-spin $M_{z}= \pm 5/2$ of the Yb ions\cite{nevidomskyy2009layered} may give rise to a vortex structure in
the hybridization between the conduction and $f$-electrons, driving a
divergent density of state at the band edges.  As the vortex line
submerges beneath the Fermi energy $E_F$, the Fermi surface undergoes a
change in topology.  Quantum criticality appears at the topological
transition where the $f$-level and its associated vortex
hybridization, are pinned at $E_F$ by charge neutrality
effects. 

Independently of topological considerations, quantum criticality without
an order parameter may arise at a quantum valence transition
\cite{watanabe10,Pixley}. While in its simplest form, this 
scenario requires an accidental, fine
tuning of a critical end point to zero temperature, 
a particularly interesting possibility is that 
the topological vortex metal could 
provide a natural way for valence
fluctuations to become  {\it quantum critical} as the $f$-level is pinned
to the Fermi energy.

Various experiments can be used to 
delineate between these scenarios.  A variety of thermodynamic measurements such as magnetization and Gr\"uneisen parameter \cite{Zhu03}, is important to confirm the strange metal phase and its quantum phase transition to the FL phase under pressure. In particular, it would be 
useful to know if the $T/B$ scaling observed in the thermodynamics
of \ybal at ambient pressure, extends throughout the region of
criticality to confirm whether the observed behavior is associated
with a critical line, forming a branch-cut in the pressure-field phase
diagram. 
Finally, it would be also useful to measure the $4f$ valence, to examine how the valence of the $4f$ state changes
in the critical pressure region. 



\bibliography{science}
\bibliographystyle{science}



We thank Y. Shimura, H. Takahashi, G. G. Lonzarich, Y. Matsumoto, E. C. T. O’Farrell, K. Matsubayashi, N. Horie, and K. Ueda for support and useful discussions. This work is partially supported by grants-in-aid (nos. 25707030 and 24740243) and the Program for Advancing Strategic International Networks to Accelerate the Circulation of Talented Researchers (no. R2604) from the Japanese Society for the Promotion of Science; by PRESTO, Japan Science and Technology Agency (JST); by Grant for Basic Science Research Projects from the Sumitomo Foundation; by U.S. National Science Foundation grant DMR-1309929 (P.C.) and U.S. National Science Foundation I2CAM International Materials Institute Award, grant DMR-0844115 (P.C. and S.N.); and by NSF grant no. PHYS-1066293 and the hospitality of the Aspen Center for Physics (P.C. and S.N.). The use of the facilities of the Materials Design and Characterization Laboratory at the Institute for Solid State Physics, The University of Tokyo, is gratefully acknowledged.


\newpage
\noindent {\bf Supplementary Materials}\\
\noindent www.sciencemag.org\\
\noindent Materials and Methods\\
\noindent Supplementary Text\\
\noindent Table S1\\
\noindent {Figs. S1 to S9}\\
\noindent References (34$\sim$ 40)

\clearpage
\newpage
\begin{center}
{\bf Figure Captions}
\end{center}
\begin{itemize}
\item[Figure 1.]  
Strange metal, Fermi liquid and magnetic order in $\beta$-YbAlB$_4$ under pressure. 
Ultrapure single crystals with the same quality (RRR = 300) were used \cite{SOM}. \textbf{(A)} Zero-field resistivity $\rho(T)$ vs. $T^{1.5}$ at various pressures (left and right axes). The anomalous $T^{1.5}$ dependence was found robust up to $P_c \sim$ 0.4 GPa. 
The superconducting (SC) transition was observed up to $P$ = 0.59 GPa. Around $P_{\rm c}$, a resistivity spike was observed just above $T_{\rm c}$ \cite{SOM}.
Inset:
$\rho(T)$ vs. $T^{1.5}$ obtained under an in-plane field $B_{ab} = 0.1$ T. The solid line indicates a fit to $T^2$ dependence found at 0.72 GPa. \textbf{(B)} $T$ dependence of the power law exponent
$\alpha= \partial \ln (\rho(T)-\rho_0)/\partial \ln T$, corresponding to $\rho(T)$ in the panel
\textbf{(A)} under $B$ = 0 (solid symbol) and under  $B_{ab} = 0.1$ T (open symbol). \textbf{(C)} Contour plot of
the exponent $\alpha$ in the $P$-$T$ phase diagram for zero field. (For $B_{ab} = 0.1$ T, see fig. S8). Red, green, and blue circles indicate the superconducting $T_c$, N\'{e}el point $T_{\rm N}$, and $T_{\rm FL}$ where $\rho(T)$ starts showing $T^2$ dependence, respectively. $T_{\rm FL}$ determined using  $\rho(T)$ under $B_{ab} = 0.1$ T is also shown as purple circles. $T_{\rm FL}$ becomes strongly suppressed nearby $P \sim P_{\rm c}$ (fig. S8). The solid line is a guide to the eye.
\textbf{(D)} $P$ dependence of the exponent $\alpha$ and the coefficient $A$ for $T^2$ dependence
of $\rho(T)$ estimated in two $T$ ranges: $90 \sim 120$ mK under zero field ($\alpha$:large red circles), and $40 \sim 60$ mK under zero field at $P > 0.8$ GPa ($\alpha$: yellow circles, $A$: closed squares) and under  $B_{ab} = 0.1$ T
at $P < 0.8$ GPa ($\alpha$: orange circles, $A$: open squares).  
To further evaluate the exponent without the ambiguity associated with the residual resistivity, we have also carried out the analysis of the resistivity exponent  (cross) using $\alpha=1+ \partial \ln (\partial \rho(T)/\partial T)/\partial \ln T$  in the $T$ range $40 \sim 80$ mK, by suppressing the SC under $B_{ab} = 0.1$ T at $P < 0.8$ GPa. All the results are found fully consistent with the above estimates.
A fit was made using a function $A = A_0 + A_1/(P - P_c)^{\beta}$, yielding $\beta = 0.8(1)$, $P_c$ = 0.40(5) GPa, $A_1$ = 0.05(1) $\mu \Omega $cmGPa/K$^2$ and $A_0$ = 0.43(1) $\mu \Omega $cm/K$^2$. \textbf{(E)} $P$ dependence of the residual resistivity $\rho_0$ under zero field (red) and  under $B_{ab} = 0.1$ T (orange). The background color (yellow, white, blue and green) in  \textbf{(D)} and  \textbf{(E)} is a guide to the eye. 

\item[Figure 2]
 Pressure-induced antiferromagnetism in \ybal.
 ({\bf A}) $T$ dependence of the in-plane resistivity $\rho(T)$ obtained under various pressures in a cubic anvil cell above $T > 2$ K. Inset: $d \rho/ dT$ vs. $T$. The kink marked by an arrow indicates the N\' eel temperature.
({\bf B}) Contour plots of the power law exponent $\alpha  = {\pd \ln (\rho(T)
- \rho_0)}/{\pd \ln T}$ of $\rho(T)$ in the $P$-$T$ phase diagram of an ultrapure
single-crystal of $\beta$-YbAlB$_4$ (RRR=$300$).  Its low $T$
and low $P$ region specified by the blue frame in panel ({\bf B}) corresponds to the one in Fig. \ref{PD}{\bf C}.  For
clarity, the values of $T_{\rm c}$ is magnified by a factor of 10. 

\item[Figure 3]
Chemical substitution effects in \ybalfe.
({\bf A}) Inelastic component $\rho(T) - \rho_0$ and ({\bf B}) the corresponding power law exponent $\alpha(T)$ vs. $T^{1.5}$ for \ybalfe with various $x$(Fe)
at ambient pressure. Inset indicates the Fe doping dependence of the residual resistivity $\rho_0$.
({\bf C}) $T$ dependence of the DC susceptibility $M/H$ (right axis) measured in both zero-field-cooling (ZFC) and field-cooling
(FC) sequences under a field of 0.1 T parallel and perpendicular to the $ab$-plane and the
magnetic part of the zero-field specific heat $C_{\rm M}$ (left axis) obtained for
$\beta$-YbAl$_{1-x}$Fe$_{x}$B$_4$ ($x = 0.06$) at ambient pressure \cite{KugaPRB}. The in-plane susceptibility for $\beta$-LuAl$_{1-x}$Fe$_{x}$B$_4$ ($x = 0.04$) is also shown. 

\item[Figure 4]
3D phase diagram of emergent electronic phases versus pressure $P$, Fe concentration
$x$, and temperature $T$ for $\beta$-YbAl$_{1-x}$Fe$_x$B$_4$. $T_{\rm c}$ and $T_{\rm FL}$ respectively denote the superconducting
transition temperature and the onset of Fermi liquid $T^2$ dependence of the
in-plane resistivity $\rho(T)$. The $P$ dependence of the N\'{e}el point $T_{\rm N}$ obtained for three
different samples with $x$(Fe) = 0, 0.02 and 0.06 is shown \cite{SOM}. For clarity, the values of $T_{\rm c}$
and $T_{\rm FL}$ are magnified by a factor of 10. The regions connecting the (non-)Fermi liquid
regions in $P$-$T$ and $x$(Fe)-$T$ phase diagrams are schematically shown in blue (yellow). Solid and broken lines are guides to the eye.

\end{itemize}

\newpage
\newpage

\newpage

\clearpage

\newpage
\renewcommand{\figurename}{Fig. S\!\!}
\setcounter{figure}{0}
\renewcommand{\tablename}{Table S\!\!}
\setcounter{table}{0}
\setcounter{page}{1}

\section*{Supplementary Materials}

\subsection*{Materials and methods}
Single crystals of $\beta$-YbAl$_{1-x}$Fe$_x$B$_4$ ($x\le 0.06$) were grown by aluminum self-flux method \cite{macaluso2007crystal,KugaPRB}.
The Fe concentration was estimated by energy dispersive X-ray analysis (EDX) within the resolution of 3\%. We also utilized the inductively coupled plasma (ICP) spectroscopy to determine $x$ (resolution of 0.3\%) for some of $\beta$-YbAl$_{1-x}$Fe$_{x}$B$_{4}$ samples and confirmed the Fe concentration within 1\% difference from the EDX results. For example, a single crystal of $\beta$-YbAl$_{1-x}$Fe$_{x}$B$_{4}$ is found to have $x = 0.03$ by EDX, and $x = 0.02$ by ICP, respectively. Throughout this paper, we use $x$(Fe) determined by EDX method for all the samples of $\beta$-YbAl$_{1-x}$Fe$_{x}$B$_{4}$.
As we will discuss in the following section, our chemical analysis using the scanning microscope indicates homogeneous distribution of Fe and Al.

\begin{figure}[t]
\begin{center}
\includegraphics[scale =0.2]{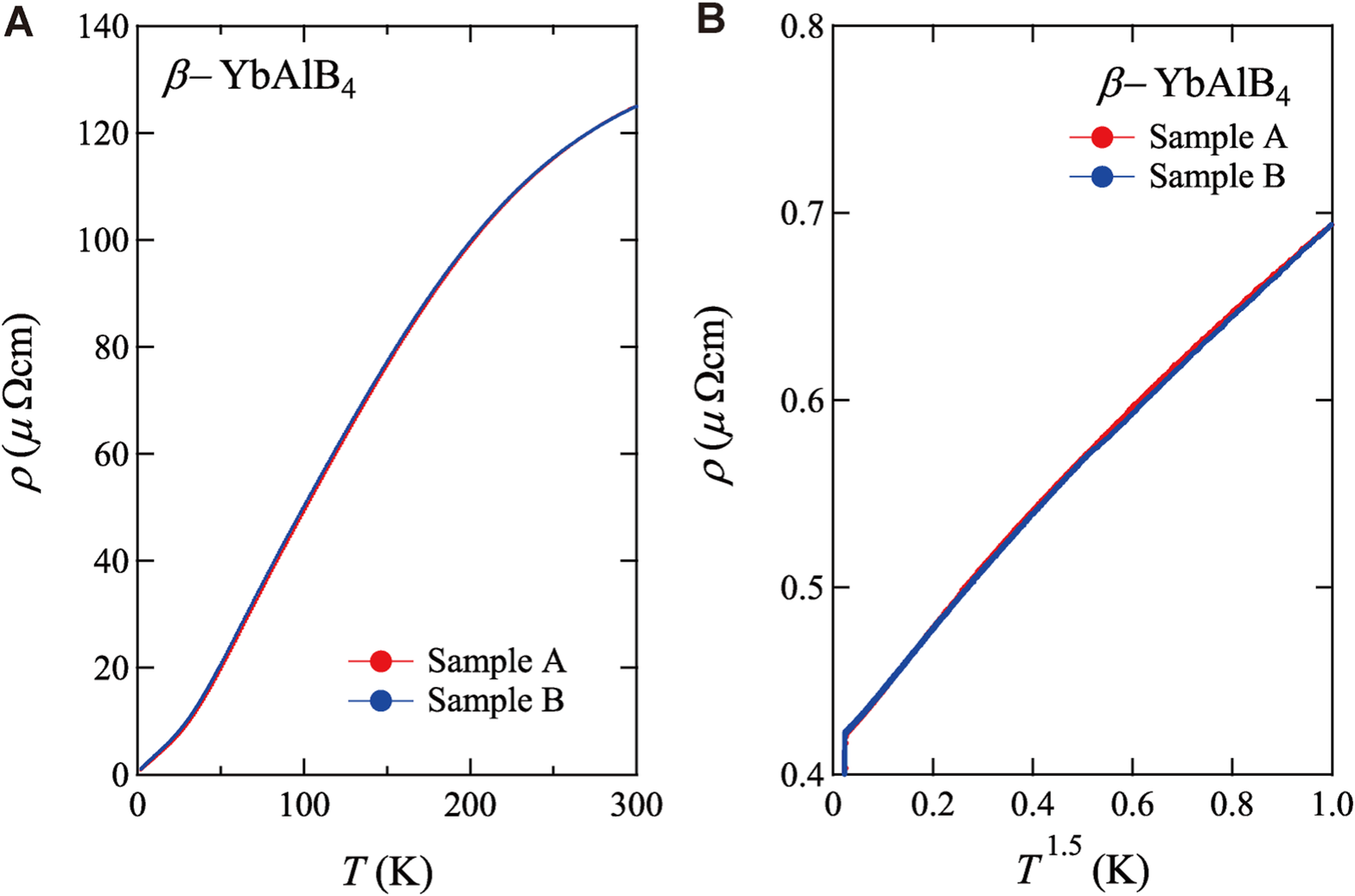}
\caption{
Temperature dependence of the in-plane electrical resistivity for two ultrapure single crystals (sample A and B) with RRR$ = 300$ employed for the low temperature resistivity measurements under pressure using a piston cylinder cell. The panel ({\bf A}) is for $T>2$ K and the panel ({\bf B}) for low temperature region at $T<1$ K.}
\label{S1}
\end{center}
\end{figure}

For resistivity measurements, ultrapure single crystals with residual resistivity ratio (RRR) more than 200 were employed.
In particular for the pressure study below 3 GPa, we used an ultrapure single crystal with RRR = 300 (with residual resistivity less than 0.5 $\mu \Omega$cm, and mean free path longer than 1000 \AA \cite{Review7, Eoin_PRL09}). We confirmed that all the samples exhibit qualitatively the same behavior. For example, Figure S\ref{S1} shows that the temperature dependence of the resistivity is almost the same for the two crystals used for the piston cylinder cell measurements.

To obtain the low temperature resistivity with high precision and accuracy, thin plate-like single crystals (typical size with 1.5 mm x 0.5 mm x 0.01 mm) were prepared.
Electrical contacts were made to the crystals by spot welding technique, yielding contact resistance of $\sim 0.1 ~ \Omega$ at room temperature.
To apply hydrostatic pressure up to $\sim 3$ GPa, these crystals were mounted in a compact (hybrid CuBe/NiCrAl) piston-cylinder-type cell together with tin and aluminium superconducting manometers, and Daphne 7373 was used as pressure transmitting medium.
A standard AC four probe method was employed using  a $^3$He/$^4$He dilution refrigerator equipped with low temperature transformers, which provide amplification of a factor of 30 and help us to reduce noise levels to less than $40$ pVHz$^{-1/2}$.
Low excitation currents ($< 30~ \mu$A) were applied below 1.5 K to avoid sample heating. To ensure thermal equilibrium between the samples and thermometers, temperature between 40 mK and 1.5 K was controlled with a slow rate  of the order of 1 mK/min.  No hysteresis in resistivity was found between warming and cooling runs.  Thermal gradient across the pressure cell was monitored by two thermometers located at both top and bottom parts of the cell and was found less than 5 mK.
To apply hydrostatic pressure up to 8 GPa in the temperature range between 2 K and 300 K, a cubic-anvil-type cell was used with pressure medium Daphne oil 7373 \cite{Review13}.
Above 2.2 GPa, there might be a slight anisotropy in pressure due to the solidification of the pressure medium at room temperature.

For $\beta$-YbAl$_{1-x}$Fe$_x$B$_4$,  we made ambient pressure measurements of electrical resistivity, DC magnetization, and specific heat.
AC-resistivity measurements were performed down to 40 mK using a $^3$He/$^4$He dilution refrigerator.
Magnetization $M$ above 2 K was measured with a commercial SQUID magnetometer.
The specific heat measurements were carried out using a thermal-relaxation technique down to 0.4 K.
The magnetic part of the specific heat $C_{\rm M}$ was
estimated by subtracting the specific heat of the nonmagnetic analog $\beta$-LuAlB$_4$.

Lattice parameters for $\beta$-YbAl$_{1-x}$Fe$_x$B$_4$ ($x\le 0.06$) were determined by single X-ray diffraction measurements using a commercial system (Rigaku Rapid II Single Crystal X-Ray Diffractometer).
On the other hand, powder X-ray diffraction measurements under pressure were carried out using a commercial X-ray diffraction system (Rigaku Micromax-007HF, $\lambda_{\mathrm{Mo}}$=0.7103 \AA).
A diamond anvil cell (DAC) with pressure transmitting medium, Daphne 7474 was used to generate pressure up to 10 GPa.
Pressure calibration was made by a ruby luminescence method.
Diffraction patterns were analyzed to determine the crystal structure using the Rietveld analysis program RIETAN-FP \cite{sp1}.




\subsection*{Supporting online text}


\subsection*{\label{sec:level1}1. Sample Properties of $\beta$-YbAl$_{1-x}$Fe$_x$B$_4$}

\begin{figure}[t]
\begin{center}
\includegraphics[scale =0.3]{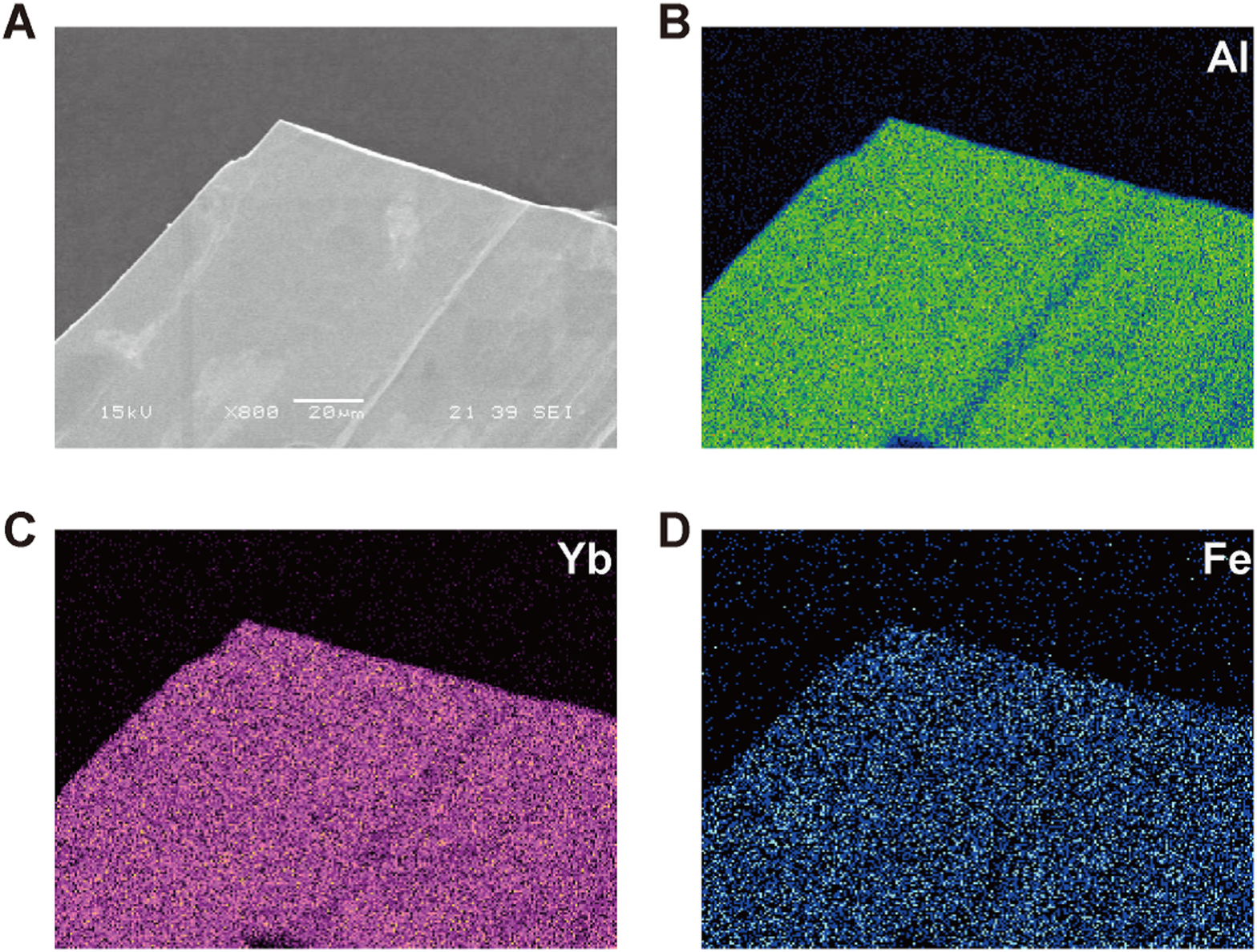}
\caption{Room temperature SEM-EDX analyses for a single crystal $\beta$-YbAl$_{1-x}$Fe$_{x}$B$_4$ ($x=0.04$).
({\bf A}) SEM image of an $ab$-plane surface and the associated  ({\bf B})Al,  ({\bf C})Yb, and  ({\bf D})Fe EDX mapping are shown.
The maps were obtained under 800 $\times$ magnification with accelerating voltage of 15 kV.
}
\label{S2}
\end{center}
\end{figure}

We estimated the Fe concentration $x$ of $\beta$-YbAl$_{1-x}$Fe$_{x}$B$_4$ using ICP and EDX methods as described above.
The chemical homogeneity was confirmed by using the SEM (scanning electron microscope)-EDX method.
Figure S\ref{S2} shows the SEM-EDX mapping of an $ab$-plane surface of a single crystalline $\beta$-YbAl$_{1-x}$Fe$_{x}$B$_4$ ($x=0.04$).  
The scanning microscope map of Fe and Al indicates homogeneous mixing of Fe and Al.
The lattice parameters and the volume of the single crystals $\beta$-YbAl$_{1-x}$Fe$_{x}$B$_4$ ($0 \le x \le 0.06$) are shown 
in Figs. S\ref{S3}{\bf A} and {\bf B} and Table S1. We found a systematic decrease of the volume as doping Fe. 
The $c$-axis compressibility is found nearly twice larger than those for the $ab$-plane.

\subsection*{\label{sec:level2}2. Crystal Structure of $\beta$-YbAl$_{1-x}$Fe$_x$B$_4$ under Pressure}
\begin{figure}[h!]
\begin{center}
\includegraphics[scale =0.2]{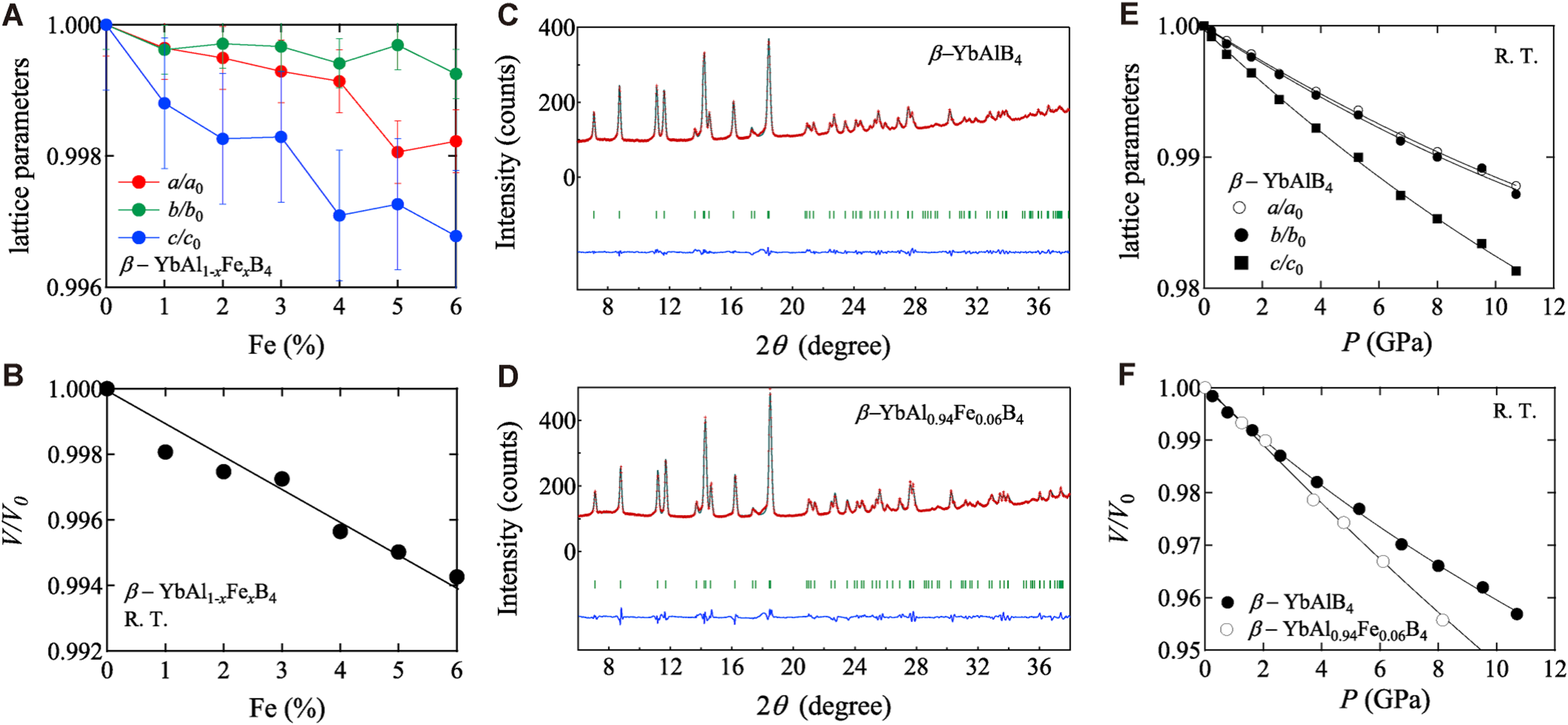}
\caption{X-ray diffraction analysis for $\beta$-YbAl$_{1-x}$Fe$_{x}$B$_4$. 
({\bf A}) Lattice parameters ($a$, $b$, and $c$)  and  ({\bf B}) unit-cell volume ($V$) as a function of the Fe-doping concentration $x$ in $\beta$-YbAl$_{1-x}$Fe$_{x}$B$_4$ at room temperature and at ambient pressure, estimated by single crystal X-ray diffraction measurements. The results are normalized by the ambient pressure values $a_0$, $b_0$, $c_0$, and $V_0$, respectively.
({\bf C \& \bf D}) Powder X-ray diffraction spectra (red crosses) are shown for ({\bf C}) $\beta$-YbAlB$_4$  and ({\bf D}) $\beta$-YbAl$_{1-x}$Fe$_{x}$B$_4$ ($x=0.06$) at room temperature and ambient pressure.
The blue line at the bottom represents the difference between the observed data (red crosses in the top) and the calculated data (dark green solid line in the top) obtained by Rietveld-refinement program RIETAN-FP \cite{sp1}.  The green bars in the middle show the positions of calculated ($hkl$) reflections.
({\bf E}) Pressure dependence of lattice constants $a$, $b$, and $c$ of $\beta$-YbAlB$_4$ at room temperature, estimated using the powder X-ray diffraction measurements. The results are normalized by the ambient pressure values $a_0$, $b_0$, and $c_0$, respectively.
({\bf F}) Pressure dependence of unit-cell volume for $\beta$-YbAl$_{1-x}$Fe$_{x}$B$_4$ ($x=0$ and 0.06)  at room temperature, normalized by the ambient pressure value $V_0$. }
\label{S3}
\end{center}
\end{figure}

Figures S\ref{S3}{\bf C} and {\bf D} show the X-ray diffraction spectra for $\beta$-YbAlB$_{4}$ and $\beta$-YbAl$_{1-x}$Fe$_{x}$B$_{4}~(x =0.06)$, respectively, at room temperature and ambient pressure.
The patterns agree well with the previously reported orthorhombic structure ($Cmmm$) for these materials \cite{macaluso2007crystal,KugaPRB}.
The results obtained by the Rietveld analysis are shown in Table S1.
The lattice constants and unit cell volume of pure $\beta$-YbAlB$_4$ are larger than those of the Fe-doped sample, indicating a chemical pressure effect induced by Fe-doping in $\beta$-YbAl$_{1-x}$Fe$_{x}$B$_{4}$ \cite{macaluso2007crystal,KugaPRB}.

Pressure dependence of the lattice constants $a$, $b$, and $c$ of pure $\beta$-YbAlB$_4$ indicates that the $c$-axis compressibility is nearly twice larger than those for the $ab$-plane (Fig. S\ref{S3}{\bf E}).
The pressure dependent unit-cell volume for both $\beta$-YbAlB$_4$ and $\beta$-YbAl$_{1-x}$Fe$_x$B$_4$  ($x=0.06$)
at room temperature are plotted in Fig. S\ref{S3}{\bf F}.
The compression of the unit cell volume can be fit with a third-order Birch-Murnaghan equation of state \cite{Birch},
$P(V)=3/2 B_0[(V_0/V)^{7/3}-(V_0/V)^{5/3}]\left\{1+3/4(B'-4)[(V_0/V)^{2/3}-1]\right\}$,
where $V$ is the unit cell volume under pressure, $V_0$ is the unit cell volume at ambient pressure, $B_0$ is the bulk modulus, and $B'$ is the first pressure derivative of the bulk modulus.
Fitting to the equation (Fig. S\ref{S3}{\bf F}, solid lines) yields the isothermal compressibility, $\kappa=(-1/V)(dV/dP)_T=5.3 \pm 0.2$ ($5.7 \pm 0.2$) $\times 10^{-3}$  GPa$^{-1}$, bulk modulus $B_0 = 1/\kappa=189 \pm 7$ ($176 \pm 6$) GPa, and first derivative of bulk modulus $B'=12 \pm 2$ ($2 \pm 1$) for $\beta$-YbAl$_{1-x}$Fe$_{x}$B$_4$ with $x=0$ (0.06).

\subsection*{\label{sec:level3}3. Specific Heat and Magnetization for $\beta$-YbAl$_{1-x}$Fe$_x$B$_4$ at Low Temperature}
\begin{figure}[h!]
\begin{center}
\includegraphics[scale =0.2]{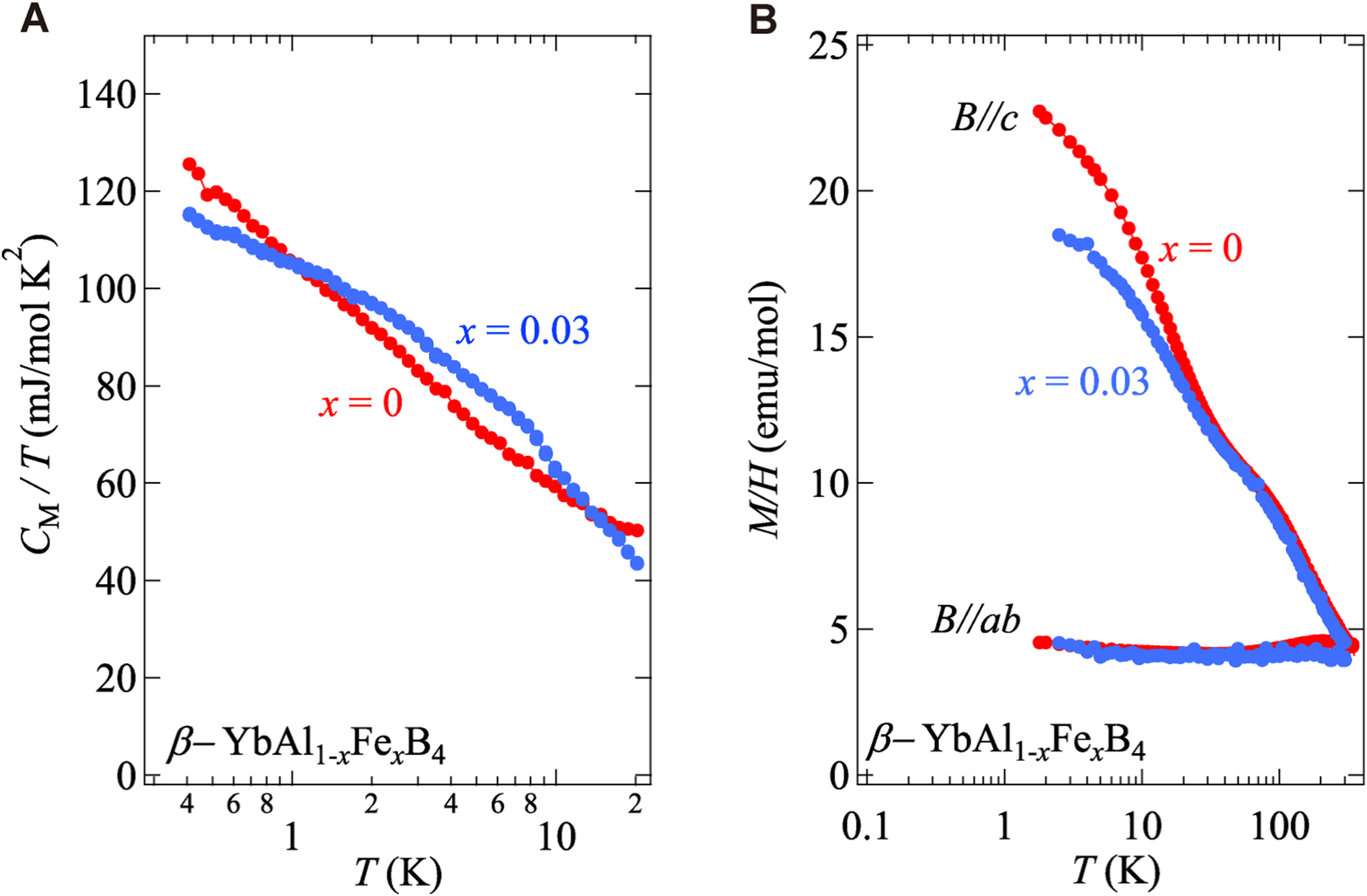}
\caption{Thermodynamic properties of $\beta$-YbAl$_{1-x}$Fe$_x$B$_4$ ($x = 0$ and 0.03).
({\bf A}) Temperature dependence of the magnetic part of the specific heat divided by temperature $C_{\rm M}/T$ of $\beta$-YbAl$_{1-x}$Fe$_x$B$_4$ ($x = 0$ and 0.03) at zero field.  $C_{\rm M}$ is estimated by subtracting the specific heat of $\beta$-LuAlB$_4$.
({\bf B}) Temperature dependence of the susceptibility $\chi \equiv M/H$ for $\beta$-YbAl$_{1-x}$Fe$_x$B$_4$ ($x =0$ and 0.03) obtained under the field of 0.1 T along the $ab$-plane and the $c$-axis. In contrast with the $c$-axis component, the $ab$-plane susceptibility is nearly $T$ independent, indicating the Ising anisotropy \cite{Review6,nevidomskyy2009layered}. No hysteresis was observed between zero-field-cooling and field-cooling sequences.
}
\label{S4}
\end{center}
\end{figure}

Figures S\ref{S4}{\bf A} and {\bf B} show the temperature dependences of the magnetic part of the specific heat divided by temperature  $C_{\rm M}/T$ and the susceptibility $M/H$ of $\beta$-YbAl$_{1-x}$Fe$_{x}$B$_{4}$ ($x=0$ and 0.03) at ambient pressure. 
$C_{\rm M}/T$ for $x=0$ shows non-Fermi liquid behavior with a logarithmic increase with decreasing $T$ \cite{Matsumoto2011Science}.
In contrast, $C_{\rm M}/T$ for the 3\% Fe-doped sample shows a weaker $T$ dependence and levels off below 10 K.
Similarly, the temperature dependence of the $c$-axis susceptibility of the 3\% Fe-doped sample shows a weaker $T$ dependence  than the pure material ($x = 0$).
These behaviors are consistent with a Fermi liquid ground state in the Fe-doped samples as indicated by $T^2$ dependence of the resistivity (Fig. 3{\bf A} in the main text).



\subsection*{\label{sec:level4}4. Superconductivity in $\beta$-YbAlB$_4$ under Pressure}



\begin{figure}[h!]
\begin{center}
\includegraphics[scale =0.2]{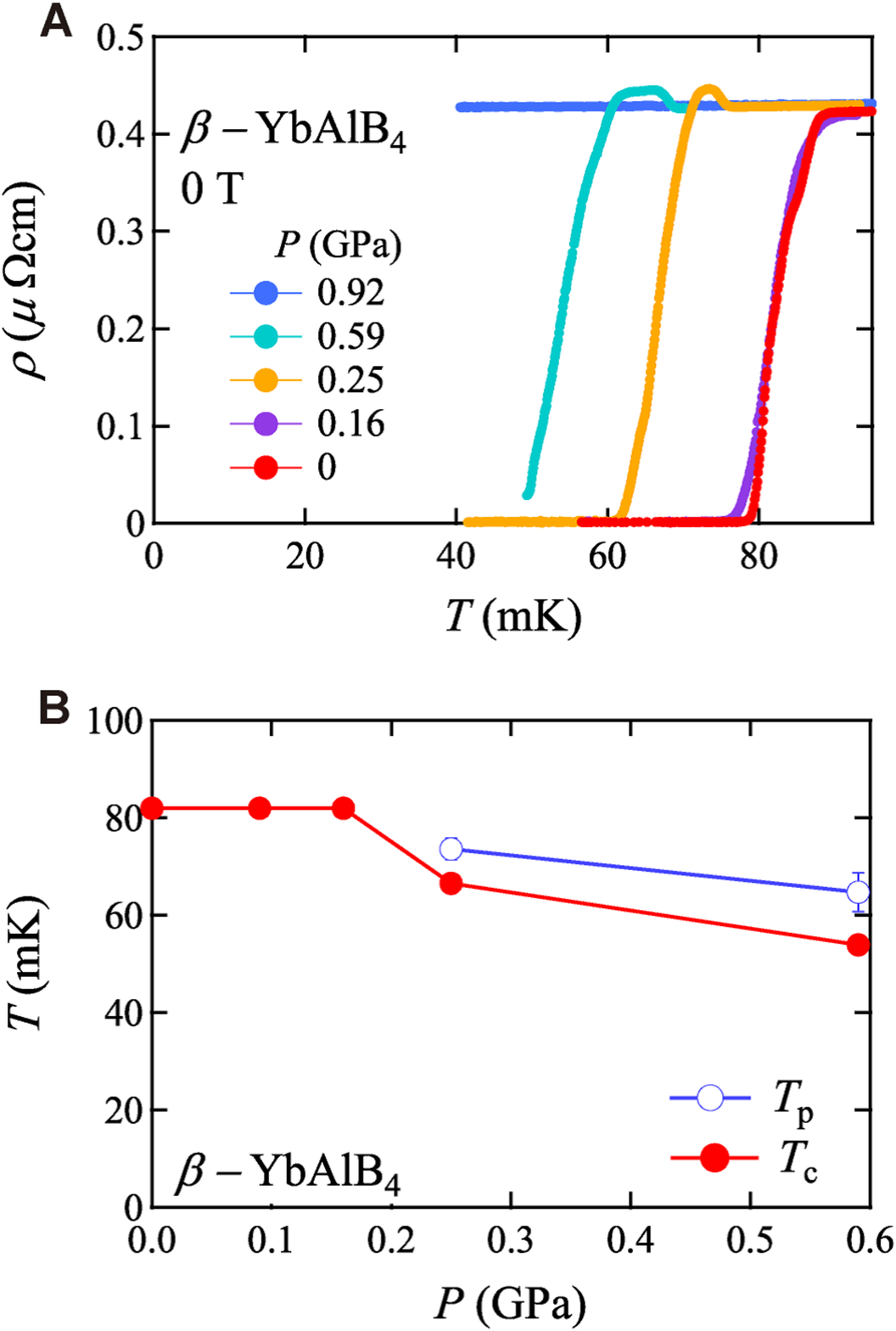}
\caption{Pressure dependence of the superconductivity of an ultrapure single crystal of \ybal with RRR = 300.
({\bf A}) Temperature dependence of the in-plane resistivity $\rho(T)$ near the superconducting transition under various pressures.
({\bf B}) Pressure dependence of the critical temperature of the superconducting transition ($T_{\rm c}$, solid circle) and the peak temperature of the resistivity ($T_{\rm p}$, open circle) observed just above $T_{\rm c}$.
}
\label{S5}
\end{center}
\end{figure}

Figure S\ref{S5}{\bf A} shows the temperature dependence of the resistivity of an ultrapure single crystal of $\beta$-YbAlB$_4$ near the superconducting transition under various pressures. The critical temperature ($T_{\rm c}$) of the superconducting transition was determined by the point where the resistivity becomes half of the normal state value at the onset of the transition. Interestingly, above $P = 0.25$ GPa, the resistivity peaks at $T_{\rm p}$ slightly above $T_{\rm c}$. This suggests formation of spatially inhomogeneous superconducting regions in the single crystal, possibly owing to a small inhomogeneity in the pressure.
Both $T_{\rm c}$ and $T_{\rm p}$ systematically decrease with pressure and become lower than the lowest $T \sim $ 40 mK of the measurements at $P=0.92$ GPa (Figs. S\ref{S5}{\bf A} and {\bf B}).

\subsection*{\label{sec:level5}5. Resistivity of $\beta$-YbAlB$_4$ under Pressure}



\begin{figure}[h!]
\begin{center}
\includegraphics[scale =0.2]{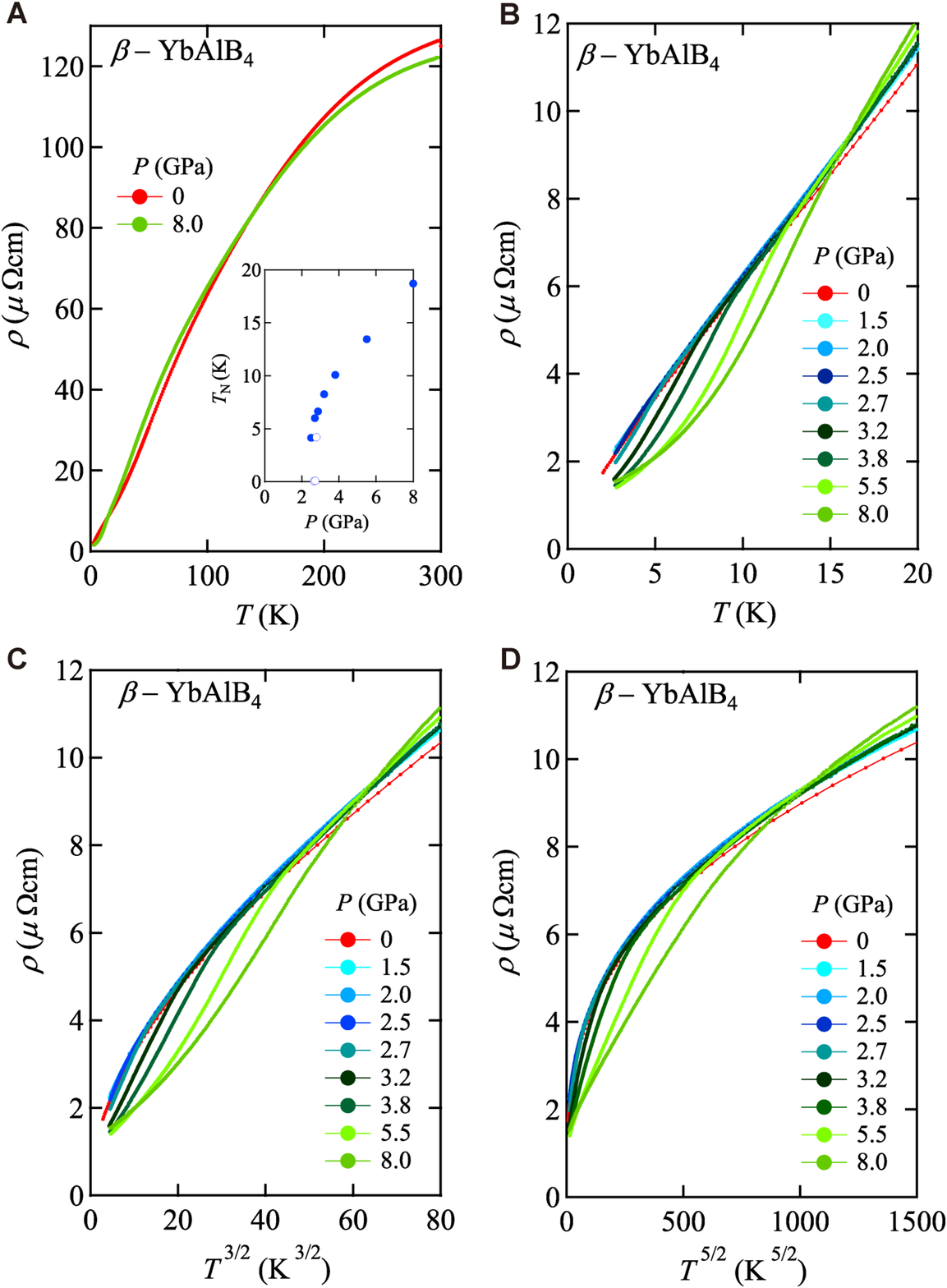}
\caption{ Temperature dependence of the in-plane resistivity $\rho(T)$  of \ybal single crystal (RRR $=200$) measured under pressure using a cubic-anvil type cell. ({\bf A}) $\rho(T)$ under ambient pressure and 8 GPa, in the temperature range between 2 and 300 K. Inset: pressure dependence of the antiferromagnetic transition temperature $T_{\rm N}$ determined by an anomaly in the temperature derivative of the resistivity $d\rho/dT$ (Fig. 2{\bf B} inset of the main text). Closed and open circles show  $T_{\rm N}$ determined by using a cubic-anvil and piston-cylinder type pressure cells, respectively.  ({\bf B}) Low temperature part of $\rho(T)$ measured under various pressures up to 8 GPa.
Panels ({\bf C}) and ({\bf D}) show $\rho(T)$ vs. $T^{3/2}$ and $T^{5/2}$, respectively.}
\label{S6}
\end{center}
\end{figure}

Figures S\ref{S6}{\bf A} and {\bf B} display the temperature dependence of the in-plane resistivity $\rho(T)$ of a high-quality single crystal of $\beta$-YbAlB$_4$ (RRR $=200$) under pressure between 0 and 8 GPa measured using a cubic anvil pressure cell.
While no anomaly in $\rho(T)$ was observed below $P= 2.1$ GPa, a pressure-induced kink appears above $P_{\rm N}\sim 2.5$ GPa.
As discussed in the main text, the kink is due to an antiferromagnetic (AF) phase transition and the loss of spin scattering should be the origin of the resistivity drop observed below the kink temperature.
The magnetic transition temperature $T_{\rm N}$ is gradually enhanced with application of pressure and reaches up to 18 K under 8 GPa  (Fig. S\ref{S6}{\bf A} inset).

A similar kink was also observed in the temperature dependence of the resistivity under pressure measured using a piston-cylinder type pressure cell in a variable temperature insert system. The transition temperatures determined by the anomaly in  the temperature derivative $d \rho/dT$ are found consistent with the phase diagram made by using the cubic anvil pressure cell (Fig. S\ref{S6}{\bf A} inset).

In the magnetically ordered phase observed above $P_{\rm c}$, the resistivity power-law exponent $\alpha$ defined by $\rho=\rho_0+ AT^{\alpha}$ gradually changes from 1 below $P \sim 2.7$ GPa, 3/2 at $3 \sim 4$ GPa, finally to 5/2 above 4 GPa (Fig. 2{\bf B} in the main text, Figs. S\ref{S6}{\bf B}, {\bf C} and {\bf D}).
The exponent $\alpha=5/2$, observed deep inside the antiferromagnetically ordered phase, is similar to the exponent observed for the antiferromagnetically ordered phase of CeCu$_2$(Si$_{1-x}$Ge$_{x}$)$_2$ \cite{Review1}.

\subsection*{\label{sec:level6}6. Resistivity of $\beta$-YbAl$_{1-x}$Fe$_{x}$B$_4$ under Pressure}
\begin{figure}[h!]
\begin{center}
\includegraphics[scale =0.2]{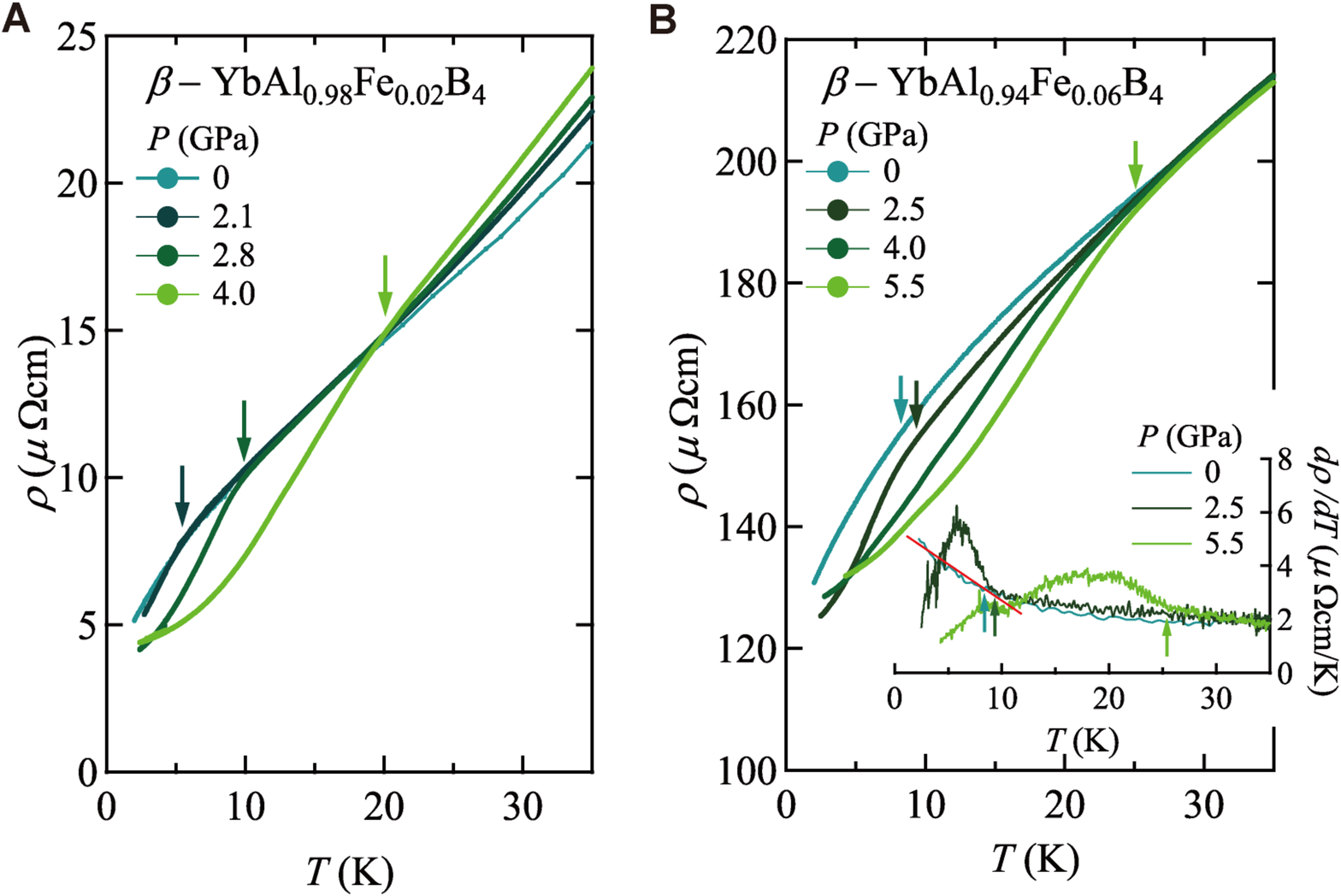}
\caption{
Temperature dependence of the in-plane resistivity $\rho(T)$ of ({\bf A}) $\beta$-YbAl$_{0.98}$Fe$_{0.02}$B$_4$ and ({\bf B}) $\beta$-YbAl$_{0.94}$Fe$_{0.06}$B$_4$ single crystals under pressure ({\bf A}) up to 4.0 GPa and ({\bf B}) up to 5.5 GPa, respectively. The pressure is applied using a cubic-anvil type pressure cell. Inset: $d \rho/ dT$
vs. $T$. The kink, marked by an arrow indicates the N\' eel temperature, $T{\rm _N}$.
}
\label{S7}
\end{center}
\end{figure}

Figures S\ref{S7}{\bf A} and {\bf B} display the temperature dependence of the in-plane resistivity $\rho(T)$ of $\beta$-YbAl$_{0.98}$Fe$_{0.02}$B$_4$ and $\beta$-YbAl$_{0.94}$Fe$_{0.06}$B$_4$ under pressure less than 5.5 GPa measured using a cubic anvil pressure cell.
The transition temperature determined by the anomaly in  the temperature derivative $d \rho/dT$ is found enhanced with pressure.
For example, as shown in the inset of Fig. S\ref{S7}{\bf B}, we found an anomaly in $d \rho/dT$ at $T{\rm _N}$ = 9 K under ambient pressure for $\beta$-YbAl$_{0.94}$Fe$_{0.06}$B$_4$, which is consistent with the magnetic ordering temperatures found in the DC magnetization and specific heat measurements (Fig. 3{\bf B} in the main text) \cite{KugaPRB}. With application of pressure, $T_{\rm N}$ estimated by the resistivity measurements is enhanced up to 26 K at 5.5 GPa.

\subsection*{\label{sec:level7}7. Low Temperature Resistivity of $\beta$-YbAlB$_4$ under Pressure}
\begin{figure}[h!]
\begin{center}
\includegraphics[scale =0.2]{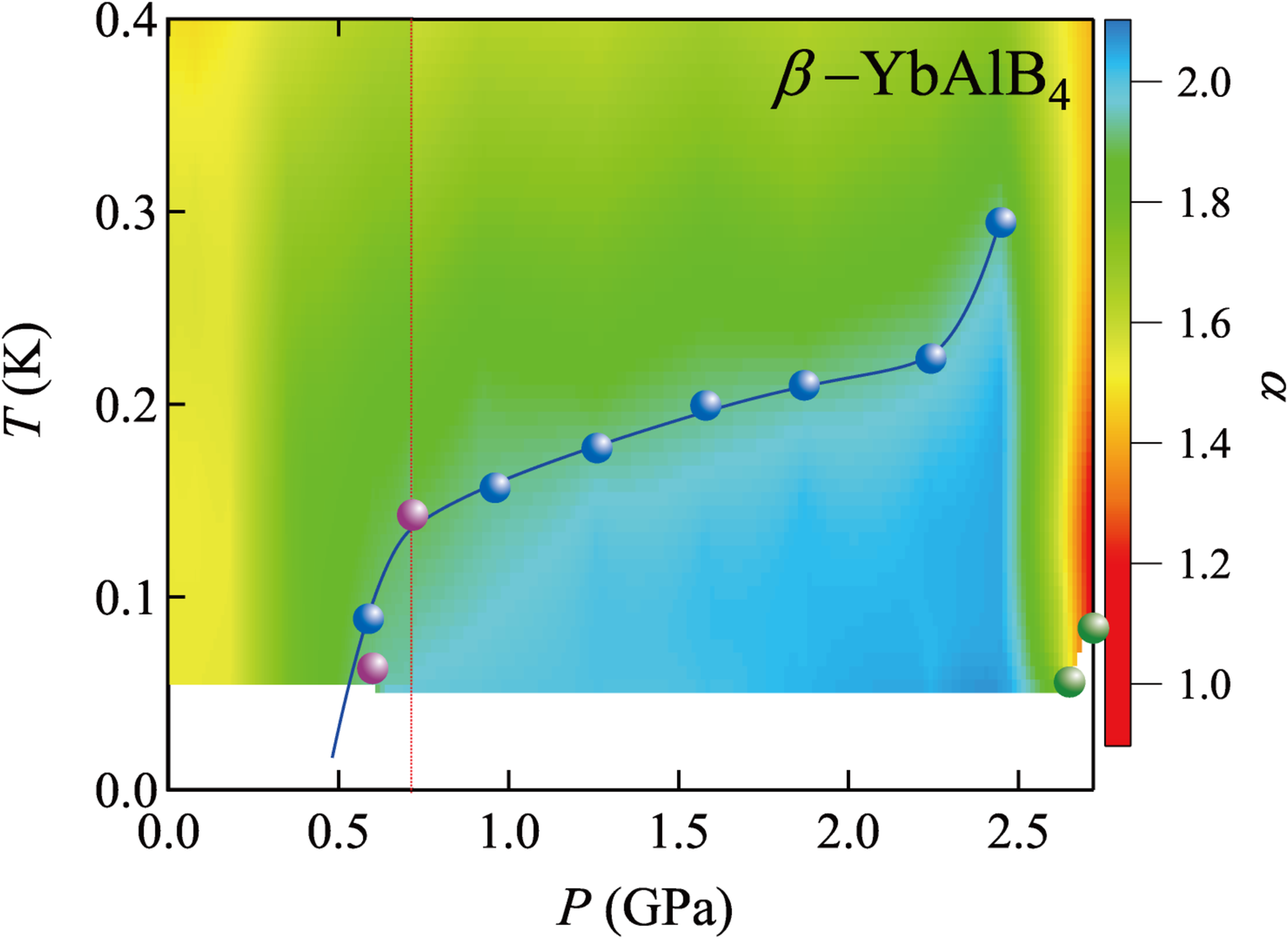}
\caption{
Contour plots of the exponent $\alpha$ of the power law behavior of the resistivity. Two data sets of the resistivity measured under the in-plane field of 0.1 T at $P \le 0.72$ GPa, and under zero field at $P > 0.72$ GPa are used and found smoothly connected with each other. The application of the field suppresses the superconductivity and allows us to reveal in detail the pressure dependence of the Fermi liquid temperature $T_{\rm FL}$ below which the resistivity shows the $T^2$ law. $T_{\rm FL}$ estimated under $B_{ab} = 0$ and $0.1$ T is shown as blue and purple circles, respectively. $T_{\rm FL}$ becomes strongly suppressed, indicating  $T_{\rm FL}$ vanishes as $P \rightarrow P_{\rm c}$. The solid blue line is a guide to the eye. The red vertical broken line indicates $P = 0.72$ GPa.}
\label{S8}
\end{center}
\end{figure}

Figure S\ref{S8} shows contour plots of the exponent $\alpha$ of the power law behavior of the low temperature resistivity. Two data sets of the resistivity measured under the in-plane field of 0.1 T at $P \le 0.72$ GPa (Fig. 1A inset) and under zero field at $P > 0.72$ GPa (Fig. 1A) are used. The high pressure part at $P > 0.72$ GPa is the same as in Figure 1C. The application of the field suppresses the superconductivity and allows us to reveal in detail the pressure dependence of the Fermi liquid temperature $T_{\rm FL}$ below which the resistivity shows the $T^2$ law. $T_{\rm FL}$ becomes strongly suppressed with decreasing pressure and appears to vanish at $P \sim P_{\rm c}$. 

\begin{figure}[h!]
\begin{center}
\includegraphics[scale =0.3]{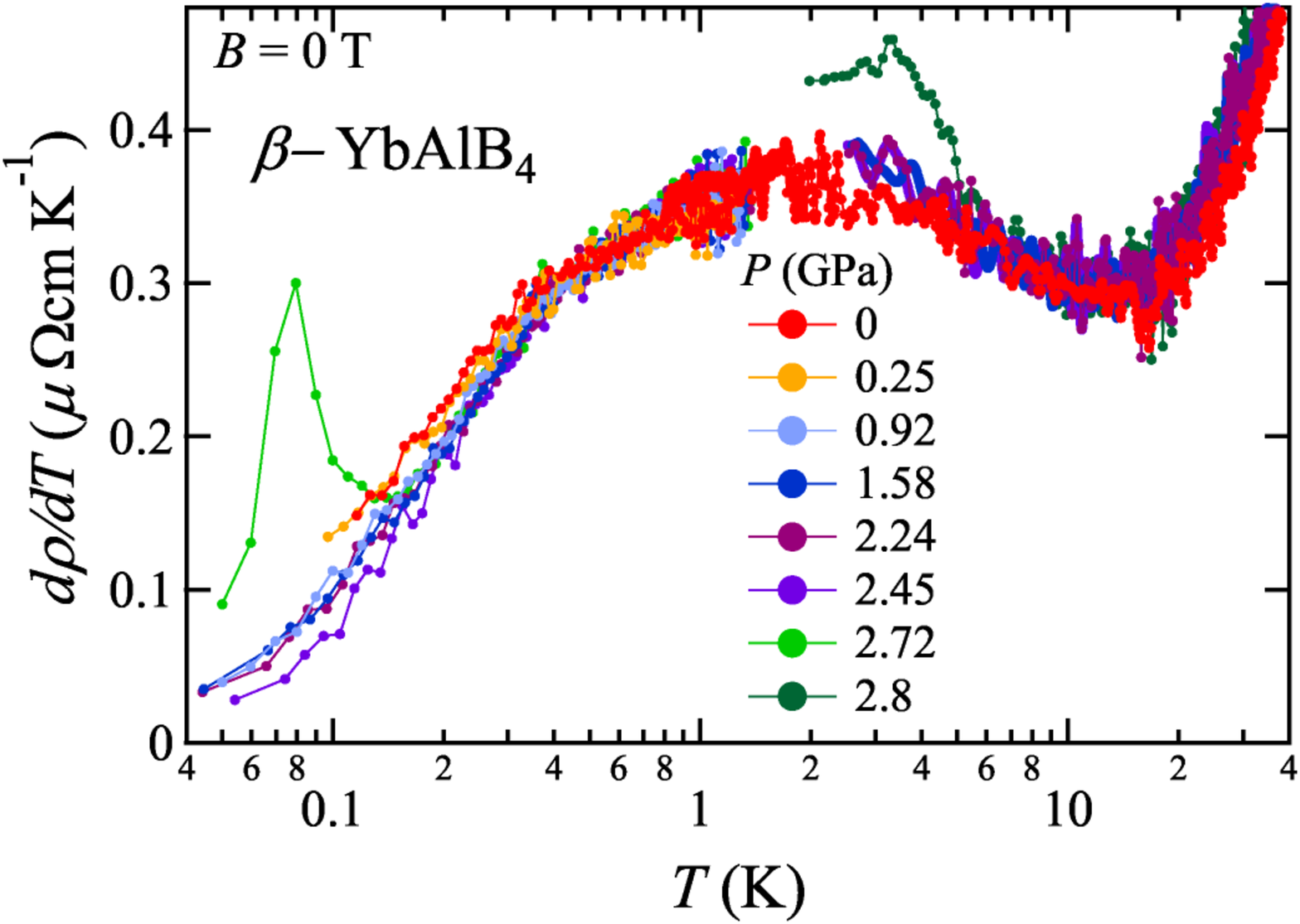}
\caption{
Temperature dependence of the temperature derivative of the zero-field resistivity, $d \rho/dT$, of $\beta$-YbAlB$_4$, measured under various pressures using a piston cylinder type cell.}
\label{S9}
\end{center}
\end{figure}

Figure S\ref{S9} shows the temperature dependence of the temperature derivative of the zero field resistivity, $d\rho/dT$,  for pure $\beta$-YbAlB$_4$, measured under various pressures using a piston cylinder type cell. All the data show a smooth change except the data for $P =$ 2.72 GPa and 2.8 GPa. This indicates the absence of magnetic order in the pressure range of $P < P_N \sim$  2.5 GPa.  At $P$ = 2.72 and 2.8 GPa ($> P_N$), a sudden increase in $d\rho/dT$ is clearly visible, indicating a magnetic order at $T_{\rm N} = 80$ mK and 4 K, respectively. The rapid growth of $T_{\rm N}$ as a function of $P$ suggests that  the pressure-induced magnetic phase transition is first order, as also indicated by the sudden change in $\rho_0$ across $P_{\rm N}$.


\newpage

\begin{table}[h!]
\caption{\label{table1}
Lattice constants and unit cell volume for $\beta$-YbAl$_{1-x}$Fe$_{x}$B$_{4}$ with various $x$ at room temperature.
The lattice parameters are estimated by single-crystal  ($^{*}$) and powder X-ray  ($^{**}$) diffraction measurements. A typical value of the residual resistivity ratio (RRR) is also shown.}
\begin{tabular}{cccccc}
&RRR &$a$ (\AA)&$b$ (\AA)&$c$ (\AA)& $V$ (\AA${^3}$)
\\
\hline
$\beta$-YbAlB$_{4} \ ^{*}$ &300 & 7.318(4) &  9.330(4) &  3.508(4) & 239.54(1)
\\
$\beta$-YbAl$_{0.99}$Fe$_{0.01}$B$_{4} \ ^{*}$ & 90 & 7.315(4) &  9.327(4) &  3.504(4) & 239.07(1)
\\
$\beta$-YbAl$_{0.98}$Fe$_{0.02}$B$_{4} \ ^{*}$ & 60 & 7.314(4) & 9.327(4) & 3.502(4) & 238.93(1)
\\
$\beta$-YbAl$_{0.97}$Fe$_{0.03}$B$_{4} \ ^{*}$ & 40 & 7.313(4) & 9.327(4) & 3.502(4) & 238.88(1)
\\
$\beta$-YbAl$_{0.96}$Fe$_{0.04}$B$_{4} \ ^{*}$ & 17 & 7.314(4) &  9.327(4) & 3.502(4) & 238.49(1)
\\
$\beta$-YbAl$_{0.95}$Fe$_{0.05}$B$_{4} \ ^{*}$&  10   & 7.304(4) &  9.327(4) & 3.499(4) & 238.34(1)
\\
$\beta$-YbAl$_{0.94}$Fe$_{0.06}$B$_{4} \ ^{*}$ & 6 & 7.305(4) &  9.323(4) &  3.497(4) & 238.16(1)
\\ \hline
$\beta$-YbAlB$_{4} \ ^{**}$ & 300 & 7.2942(4) &  9.3028(6) &  3.4952(2) & 237.174(25)
\\
$\beta$-YbAl$_{0.94}$Fe$_{0.06}$B$_{4} \ ^{**}$ &  6 & 7.2837(6) &  9.2978(8) &  3.4830(3) & 235.881(35)
\\	\hline
\end{tabular}
\end{table}

\newpage

\newpage

\newpage

\newpage

\newpage

\newpage

\newpage


\newpage

\newpage

\end{document}